\documentclass[12pt,a4paper,final]{iopart}

\usepackage{iopams}  
\usepackage{iopams}  
\usepackage{graphicx}
\usepackage[breaklinks=true,colorlinks=true,linkcolor=blue,urlcolor=blue,citecolor=blue]{hyperref}
\usepackage[usenames,dvipsnames,svgnames,table]{xcolor}

\begin{document}

\title{Effects of stoichiometric doping in superconducting Bi-O-S compounds}
\author{Corentin Morice}
\ead{cm712@cam.ac.uk}
\address{Cavendish Laboratory, University of Cambridge, Cambridge CB3 0HE, United Kingdom}
\author{Emilio Artacho}
\ead{ea245@cam.ac.uk}
\address{Cavendish Laboratory, University of Cambridge, Cambridge CB3 0HE, United Kingdom}
\address{Nanogune and DIPC, Tolosa Hiribidea 76, 20018 San Sebasti\'an, Spain}
\address{Basque Foundation for Science, Ikerbasque, 48011 Bilbao, Spain}
\author{Sian E. Dutton}
\address{Cavendish Laboratory, University of Cambridge, Cambridge CB3 0HE, United Kingdom}
\author{Daniel Molnar}
\address{Cavendish Laboratory, University of Cambridge, Cambridge CB3 0HE, United Kingdom}
\author{Hyeong-Jin Kim}
\address{Cavendish Laboratory, University of Cambridge, Cambridge CB3 0HE, United Kingdom}
\author{Siddharth S. Saxena}
\ead{sss21@cam.ac.uk}
\address{Cavendish Laboratory, University of Cambridge, Cambridge CB3 0HE, United Kingdom}

\date{\today}

\begin{abstract}
Newly discovered Bi-O-S compounds remain an enigma in attempts to understand their electronic properties. A recent study of Bi$_{4}$O$_{4}$S$_{3}$ has shown it to be a mixture of two phases, Bi$_{2}$OS$_{2}$ and Bi$_{3}$O$_{2}$S$_{3}$, the latter being superconducting [W. A. Phelan et al., J. Am. Chem. Soc. 135, 5372 (2013)]. Using density functional theory, we explore the electronic structure of both the phases and the effect of the introduction of stacking faults. Our results demonstrate that the S$_{2}$ layers dope the bismuth-sulphur bands. The bands at the Fermi level are of clear two-dimensional character. One band manifold is confined to the two adjacent, square-lattice bismuth-sulphur planes, a second manifold is confined to the square lattice of sulphur dimers. We show that the introduction of defects in the stacking does not influence the electronic structure. Finally, we also show that spin-orbit coupling does not have any significant effect on the states close to the Fermi level at the energy scale considered.
\end{abstract}

\maketitle

\section{Introduction}

The recent discovery of superconductivity in two new compounds, Bi$_{4}$O$_{4}$S$_{3}$ \cite{Mizuguchi2012,Singh2012,Takatsu2012,Tan2012} and LaO$_{x}$F$_{1-x}$BiS$_{2}$ \cite{Awana2013,Deguchi2013,Kotegawa2012,Mizuguchi2012a}, has raised great interest. All those compounds share the same structure: a stacking of alternating BiS$_{2}$ bilayers and spacer layers. Superconductivity is thought to arise from doping of the BiS$_{2}$ bilayers. Electron doping in Sr$_{1-x}$La$_{x}$FBiS$_{2}$ \cite{Lin2013}, and La$_{1-x}$M$_{x}$OBiS$_{2}$ (M = Ti, Zr, Hf, Th) \cite{Yazici2013a} results in superconductivity. The versatility of the La(O,F)BiS$_{2}$ system has been demonstrated by replacement of La with a range of lanthanide \textit{Ln}$^{3+}$ ions \cite{Demura2013,Jha2013a,Jha2013,Xing2012,Yazici2013}--- a maximum $T_{c}$ of 10.6 K is reported in LaO$_{x}$F$_{1-x}$BiS$_{2}$ for $x=0.5$.

\begin{figure}[htb]
\centering
\includegraphics[width=8cm]{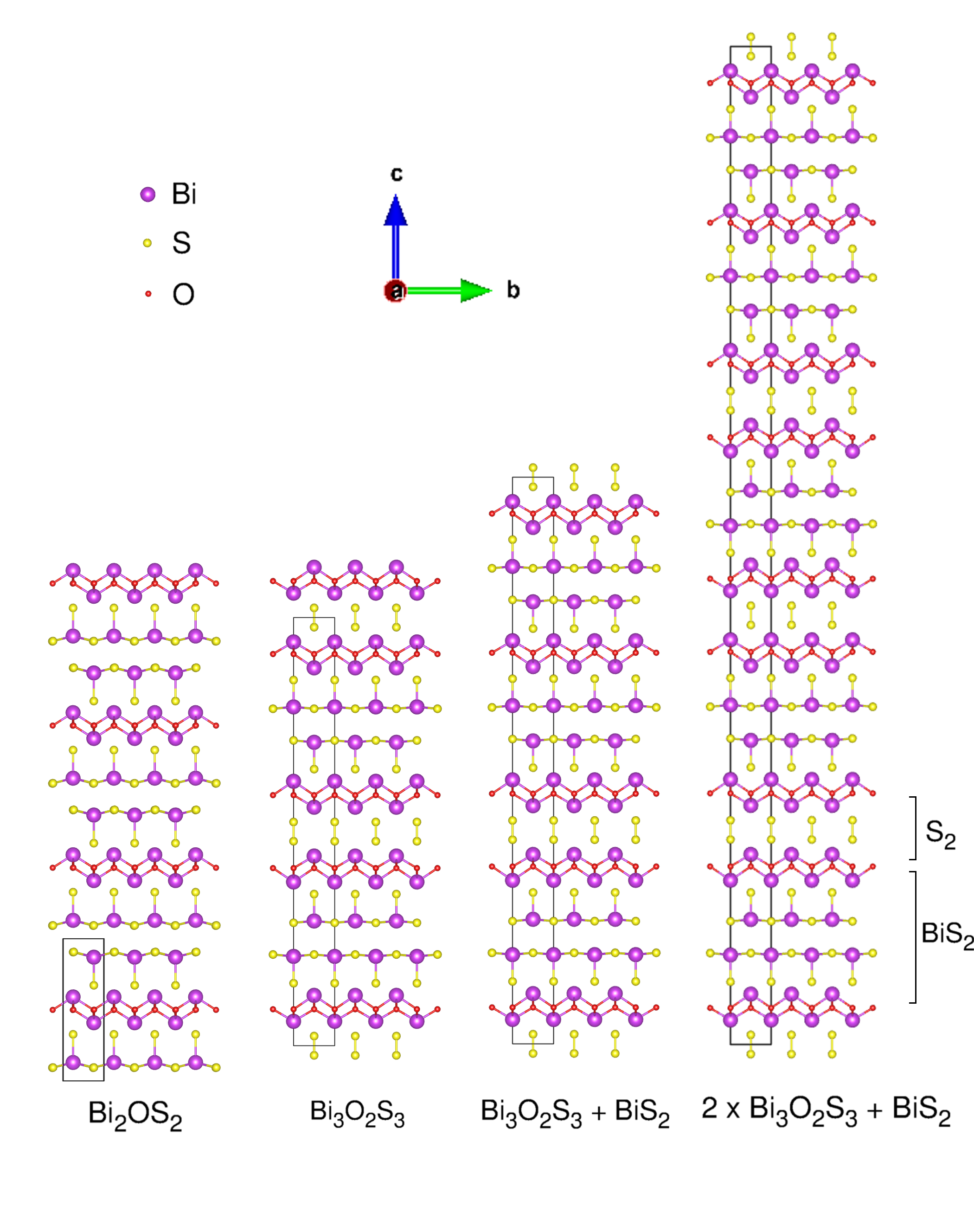}
\caption{Crystal structure of Bi$_{2}$OS$_{2}$, Bi$_{3}$O$_{2}$S$_{3}$, Bi$_{3}$O$_{2}$S$_{3}$ with one stacking fault per unit cell of Bi$_{3}$O$_{2}$S$_{3}$, and Bi$_{3}$O$_{2}$S$_{3}$ with one stacking fault per two unit cell of Bi$_{3}$O$_{2}$S$_{3}$, represented with Vesta \cite{Momma2011}. They are formed of a stacking of BiS$_{2}$, S$_{2}$ and Bi$_{2}$O$_{2}$ layers, the latter being a spacer layer. These structures implement four different S$_{2}$/BiS$_{2}$ ratios : 0, 1, 2/3, 4/5, i.e. four different frequencies of occurrence of stacking faults. The black squares represent the unit cells used in the calculations.}
\label{fig:crystal structures}
\end{figure}

Electronic structure calculations have mainly focused on the LaO$_{x}$F$_{1-x}$BiS$_{2}$ (x = 0, 0.5) materials partially due to uncertainties in the composition of the Bi-O-S superconducting phase \cite{Shein2013,Usui2012}. These calculations indicated that the superconducting electrons are a mixture of Bismuth 6\textit{p$_{x,y}$} and Sulphur 3\textit{p$_{x,y}$} states \cite{Shein2013,Usui2012}. These form 8 bands, four of which sit under the Fermi level, when the other four either are above the Fermi level or cross it. The electronic and thermodynamic properties of this group of materials are indeed exciting and enigmatic and suggestions range from spin-fluctuation mediated superconductivity \cite{Martins2013} to proximity to ferroelectricity and charge density wave (CDW) instabilities \cite{Yildirim2013,Wan2013}.

The coupling mechanism has been investigated in various ways. Electron-phonon interactions have been calculated in La(O,F)BiS$_{2}$, and yield a large electron-phonon coupling constant, suggesting superconductivity in this compound is strongly coupled and conventional \cite{Wan2013, Yildirim2013, Li2013}.  Renormalisation-group calculations suggested triplet pairing and weak topological superconductivity \cite{Yao2013,Yang2013}, a possibility studied in the context of quasiparticle interference \cite{Gao2014}. Random phase approximation was applied to a two-orbital model \cite{Usui2012}, leading to an extended s-wave or d-wave pairing \cite{Martins2013,Zhou2013}.

The first superconductor of this family to be discovered, Bi$_{4}$O$_{4}$S$_{3}$, was studied in great detail \cite{Srivatsava2013,I.Sathish2013,Shruti2013,Li2013a,Biswas2013,Selvan2013,Jha2014}. Nevertheless, there were still doubts concerning its structure. However Phelan et al. recently published an extensive study of the chemistry of these compounds, and concluded that “Bi$_{4}$O$_{4}$S$_{3}$” was actually a two-phased material \cite{Phelan2013}. The two phases are Bi$_{2}$OS$_{2}$ and Bi$_{3}$O$_{2}$S$_{3}$, the latter being assigned as the superconducting one. Single crystals of the superconducting phase have recently been synthesised \cite{Shao2014}.

\begin{figure}[htb]
\centering
\includegraphics[width=8cm]{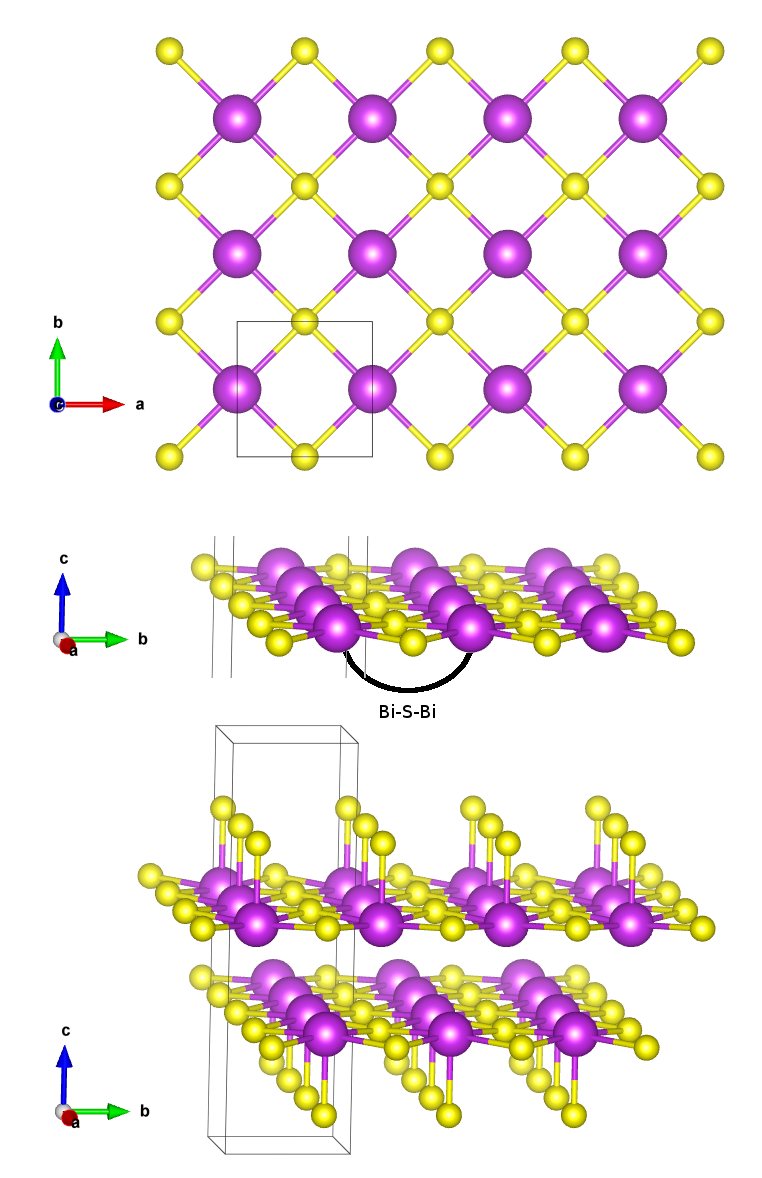}
\caption{BiS plane seen from above (top), in perspective (middle), and BiS$_{2}$ bilayer seen in perspective (bottom) represented with Vesta \cite{Momma2011}. One BiS$_{2}$ bilayer contains two BiS planes, and one extra S atom on top (or below) each Bi atom. The Bi-S-Bi angle is highlighted in the middle plot.}
\label{fig:BiS plane}
\end{figure}

Composed of BiS$_{2}$, S$_{2}$ and Bi$_{2}$O$_{2}$ layers, the latter being a “spacer” layer (Figure \ref{fig:crystal structures}), Bi$_{3}$O$_{2}$S$_{3}$ is made of alternating BiS$_{2}$ bilayers and S$_{2}$ layers, all separated by Bi$_{2}$O$_{2}$ spacer layers, whereas in Bi$_{2}$OS$_{2}$ only the two BiS$_{2}$ bilayers are present. Here, we call \textit{BiS plane} a two-dimensional squared lattice containing bismuth and sulphur atoms. We call \textit{BiS$_{2}$ bilayer} a structure containing two BiS planes and extra sulphurs localised on top or under each bismuth atom of the BiS planes. Figure \ref{fig:BiS plane} illustrates the difference between BiS plane and BiS$_{2}$ bilayer.

Experimental results \cite{Phelan2013} suggest that superconductivity is suppressed by the inclusion of additional BiS$_{2}$ bilayers disrupting the alternation of S$_{2}$ and BiS$_{2}$ sheets, which we call stacking faults. Phelan et al. studied these stacking faults in details, in particular displaying TEM data that shows without ambiguity the variations in stacking. Stacking faults play a significant role in superconducting properties of several families of materials \cite{Weger1971}. For example, changes to the ground state have been observed in the presence of stacking faults in other systems, including modifications to superconductivity \cite{Banno2012,Essmann1973}.

In this paper we explore the electronic structure of the recently experimentally determined superconducting and non-superconducting phases of the Bi-O-S compounds. The role of introducing additional BiS$_{2}$ bilayers in the superconductor Bi$_{3}$O$_{2}$S$_{3}$ is also investigated, as suggested by Phelan et al. in their experimental work \cite{Phelan2013}. Like in the La(O,F)BiS$_{2}$ systems we find that the BiS$_{2}$ bilayers play a critical role in conduction. Our analysis indicates that the S$_{2}$ sheets give rise to electron carriers in the bismuth-sulphur bands. These electrons are localised within the BiS planes rather that through the BiS$_{2}$ bilayer. Interruption of the BiS$_{2}$, S$_{2}$ stacking sequence is shown not to influence the electronic structure.

\section{Methods}

Band structures and density of states for Bi$_{2}$OS$_{2}$ and Bi$_{3}$O$_{2}$S$_{3}$ were calculated, using the SIESTA method \cite{Artacho2008,Soler2002}, implementing the generalized gradient approximation (GGA) in the shape of the Perdew, Burke, and Ernzerhof (PBE) functional. It uses norm-conserving pseudopotentials to replace the core electrons, while the valence electrons are described using atomic-like orbitals as basis states at the double zeta polarized level. In the case of the two known structures, experimentally determined structural parameters were used in our calculations \cite{Phelan2013}\footnote{For the first phase, Bi$_{2}$OS$_{2}$, the space group is P4/nmm, and the lattice parameters are $a=3.9661$  \AA\  and $c=13.798$  \AA\ . The atoms coordinates are Bi1 : (0.25,0.25,0.59122), Bi2 : (0.25,0.25,0.12989), O : (0.75,0.25,0.5), S1 : (0.25,0.25,0.328), S2 : (0.25,0.25,0.9063). For the second phase, Bi$_{3}$O$_{2}$S$_{3}$, the space group is I4/mmm, and the lattice parameters are $a=3.96721$  \AA\ and $c=41.2847$ \AA. The atoms coordinates are Bi1 : (0,0,0.05699), Bi2: (0,0,0.20865), Bi3 : (0,0,0.38245), O : (0,0.5,0.0852), S1 : (0,0,0.14595), S2 : (0,0,0.2870), S3 : (0,0,0.47748). All the occupancies are equal to one.} Spin polarized calculations were found to be unnecessary. We used a k-point grid of 16x16x5 for Bi$_{2}$OS$_{2}$, of  16x16x2 for Bi$_{3}$O$_{2}$S$_{3}$ and of 16x16x1 for the two other structures, after showing the electronic structure was well converged using them. We used an electronic temperature of 7 K and a real space mesh cutoff of 300 Ry for the real-space integration, necessary to calculate the matrix elements for some terms of the Hamiltonian. The convergence of these parameters was also checked.

In order to estimate the influence of spin-orbit coupling, we also performed all electron calculations on Bi$_{3}$O$_{2}$S$_{3}$ using full-potential linearised augmented plane waves, including spin-orbit coupling through a second-variational scheme, as implemented in the Elk program \cite{Elk}. We used the PBE functional and the same k-point grid we used with SIESTA, along with a carefully converged high-quality set of parameters.

\section{Crystal structure}

The structures of the key materials for this study, Bi$_{2}$OS$_{2}$ and Bi$_{3}$O$_{2}$S$_{3}$, are taken from the experiments in \cite{Phelan2013}. In order to simulate how stacking faults perturb the electronic structure, we also investigated materials where the relative number of BiS$_{2}$ bilayers and S$_{2}$ layers was changed, as suggested by experiment, in particular by TEM data \cite{Phelan2013}. We did not use relaxed structures in order to keep consistency with the structures directly measured with X-rays.

In Bi$_{3}$O$_{2}$S$_{3}$, the stacking is  2 x BiS$_{2}$ / Bi$_{2}$O$_{2}$ / S$_{2}$ / Bi$_{2}$O$_{2}$ / 2 x BiS$_{2}$ / Bi$_{2}$O$_{2}$ / S$_{2}$ / Bi$_{2}$O$_{2}$ (Figure \ref{fig:crystal structures}). The sequence is repeated twice because of symmetry: S$_{2}$ layers shift the Bi$_{2}$O$_{2}$ layers next to them on the x and y axis. Therefore the unit cell has to be doubled so that the stacking of one cell on another is possible.

To modify the ratio of S$_{2}$ layers over BiS$_{2}$ bilayers, we added a BiS$_{2}$ bilayer, with its corresponding Bi$_{2}$O$_{2}$ spacer layer, at the top of Bi$_{3}$O$_{2}$S$_{3}$. We therefore obtain the following stacking : 2 x BiS$_{2}$ / Bi$_{2}$O$_{2}$ / S$_{2}$ / Bi$_{2}$O$_{2}$ / 2 x BiS$_{2}$ / Bi$_{2}$O$_{2}$ / S$_{2}$ / Bi$_{2}$O$_{2}$ / 2 x BiS$_{2}$ / Bi$_{2}$O$_{2}$. This can be done because BiS$_{2}$ bilayers, unlike S$_{2}$, do not shift Bi$_{2}$O$_{2}$ layers laterally. The stacking sequence is therefore not disturbed by the added block. Similarly we also constructed a structure in which an additional BiS$_{2}$ bilayer was included every two units cells. In total, we calculated the electronic structure of compounds with four different S$_{2}$/BiS$_{2}$ ratios: 0,1, 2/3 and 4/5.

The structure of BiS$_{2}$ bilayers is different in Bi$_{2}$OS$_{2}$ and Bi$_{3}$O$_{2}$S$_{3}$, the most noticeable difference being the Bi-S-Bi angle in the BiS planes (Figure \ref{fig:BiS plane}), which is larger in the parent phase. We thus have had to choose which BiS$_{2}$ bilayer to add in the compounds with a modified stacking: Bi$_{3}$O$_{2}$S$_{3}$+BiS$_{2}$ and 2xBi$_{3}$O$_{2}$S$_{3}$+BiS$_{2}$ (Figure \ref{fig:crystal structures}). We did calculations for both, and the results were extremely similar (e.g. the bands are less than 0.1 eV apart). The results discussed below are the ones for Bi$_{3}$O$_{2}$S$_{3}$'s BiS$_{2}$ bilayers, as the stacking faults appear in this compound.

\section{Electronic structure}

\begin{figure}[htb]
\centering
\includegraphics[width=8cm]{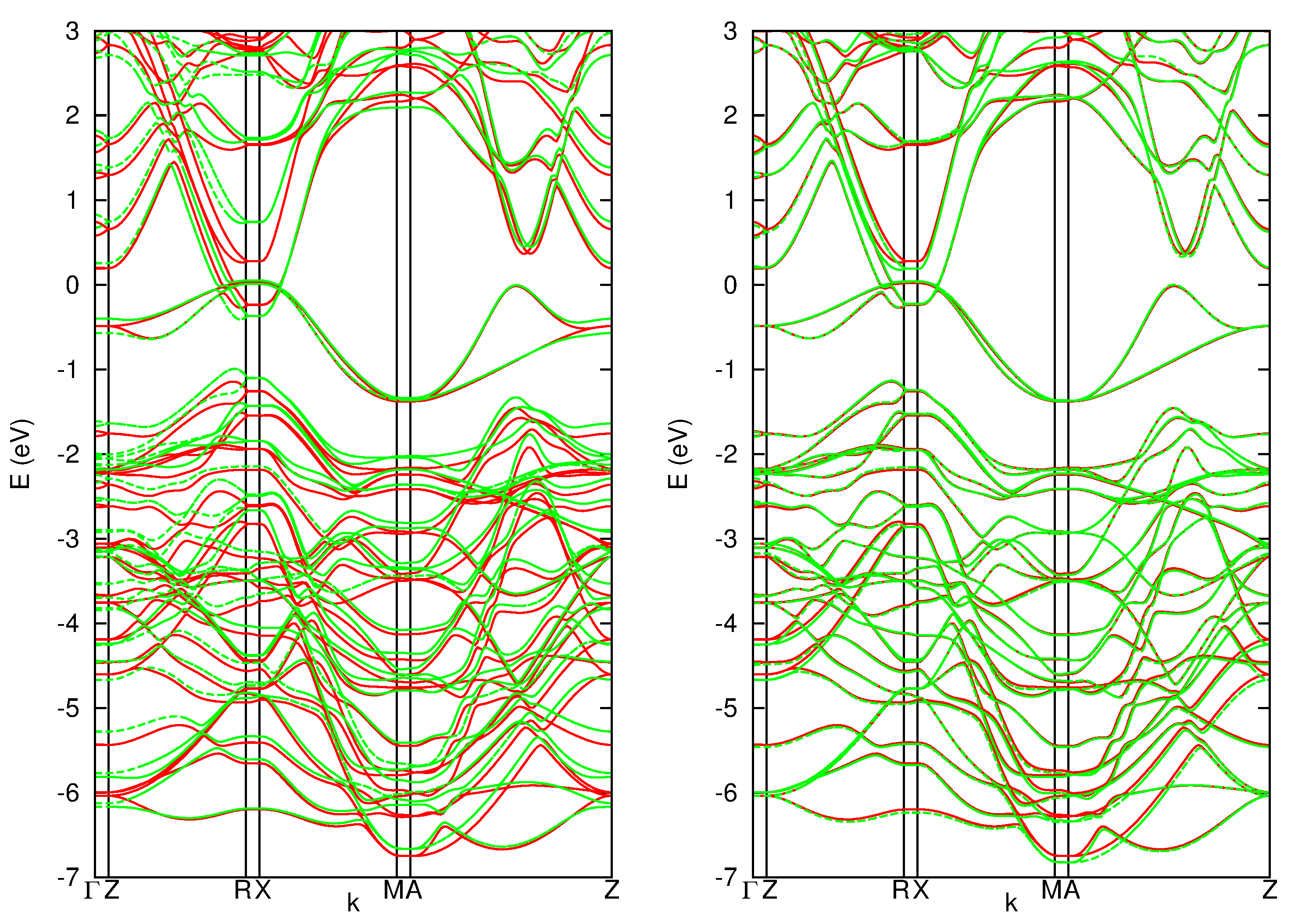}
\caption{Left: band structure of Bi$_{3}$O$_{2}$S$_{3}$ calculated with Elk, with (red) and without (green) spin-orbit coupling. Right: band structure of Bi$_{3}$O$_{2}$S$_{3}$ calculated with Elk (red) and with SIESTA (green) without spin-orbit coupling. In all band structures presented in this paper, the Fermi level is at zero energy.}
\label{fig:Elk SO comparison}
\end{figure}

The band structures calculated with Elk and SIESTA are in very close agreement (Figure \ref{fig:Elk SO comparison}, right hand side). The influence of spin-orbit coupling is very limited, with a difference under 0.15 eV between energy levels with and without spin-orbit coupling (Figure \ref{fig:Elk SO comparison}, left hand side). The presence or absence of spin-orbit coupling has no impact on any of the features of the electronic structure discussed below. In the following we concentrate on electronic structures calculated with SIESTA.

\subsection{Band structures}

\begin{figure}[htb]
\centering
\includegraphics[width=5.cm]{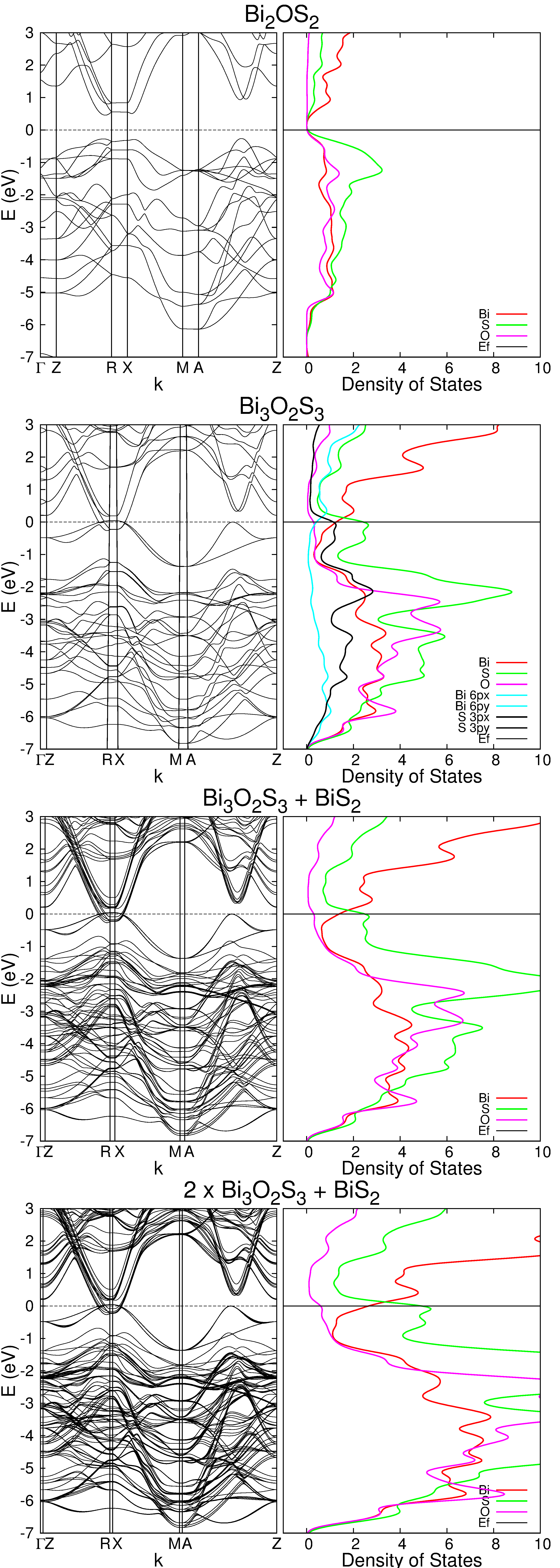}
\caption{Band structures and densities of states projected onto the basis orbitals for Bi (red), S (green) and O (lilac) for the four simulated compounds. From top to bottom, they correspond to the structures from left to right in Figure \ref{fig:crystal structures}. The two last compounds are simulations of stacking faults for two different frequencies in Bi$_{3}$O$_{2}$S$_{3}$. In Bi$_{3}$O$_{2}$S$_{3}$ without stacking faults, we plot the partial density of states for the bismuth 6\textit{p$_{x,y}$} (light blue) and sulphur 3\textit{p$_{x,y}$} orbitals (black). In each case, the \textit{p$_{x}$} and \textit{p$_{y}$} orbitals are indistinguishable. At the Fermi level, they each contribute approximately half of the density of state of the corresponding specie. These projections have been omitted in the other plots for clarity. The y-axis has been chosen so that the Fermi level is at zero energy.}
\label{fig:band structures}
\end{figure}

Bi$_{2}$OS$_{2}$ and Bi$_{3}$O$_{2}$S$_{3}$ only differ by the replacement in the second compound of one out of two BiS$_{2}$ bilayer by an S$_{2}$ layer. In terms of stoichiometry, as noted by Phelan et al. \cite{Phelan2013}, Bi$_{2}$OS$_{2}$ can be written BiOBiS$_{2}$, just as the parent phase of the lanthanum compound, LaOBiS$_{2}$.

The band structure of Bi$_{2}$OS$_{2}$ (Figure \ref{fig:band structures}) has features close to the Fermi energy which are similar to those of the undoped LaOBiS$_{2}$ phase \cite{Shein2013,Usui2012}. For all band structures calculated, of specific interest is the set of bands crossing or just above the Fermi level near the R and X points, and approaching the Fermi level between the A and Z points (Figure \ref{fig:band structures}). These were also found in LaO$_{x}$F$_{1-x}$BiS$_{2}$ \cite{Shein2013,Usui2012}, and were shown in these compounds to be composed of bismuth 6\textit{p$_{x,y}$} and sulphur 3\textit{p$_{x,y}$} orbitals. Partial density of states calculations enabled us to confirm that in all the compounds considered in this study these bands are composed of bismuth 6\textit{p$_{x,y}$} and sulphur 3\textit{p$_{x,y}$} orbitals, confirming the assignment done in \cite{Usui2012} (Figure \ref{fig:band structures}).

In the first case, Bi$_{2}$OS$_{2}$, these BiS bands are just above the Fermi level, at around 0.3 eV at their minimum. This is very close to the corresponding minimum in LaOBiS$_{2}$ (0.2 eV) \cite{Shein2013,Usui2012}. The gap is 0.7 eV wide, twice as large as in the LaOBiS$_{2}$ case (0.37 eV) \cite{Shein2013,Usui2012}.

In Bi$_{3}$O$_{2}$S$_{3}$, one of these BiS bands crosses the Fermi level near the R and X points, just as in LaO$_{0.5}$F$_{0.5}$BiS$_{2}$ \cite{Shein2013,Usui2012}. At these points it interacts with two other bands. These are in great majority composed of S 3\textit{p} states, as the projected density of states shows ("projected" refers to decomposing the total density of states in contributions from the different basis orbitals of the different atoms). These bands are not present in the parent phase, Bi$_{2}$OS$_{2}$ ; they come from the S$_{2}$ layer. These two bands therefore correspond to the partly filled $\pi$ antibonding states in the S$_{2}$ dimers. If fully occupied, this would formally correspond to S$_{2}^{2-}$ dimers.

\subsection{Local densities of states}

\begin{figure}[htb]
\centering
\includegraphics[width=5.5cm]{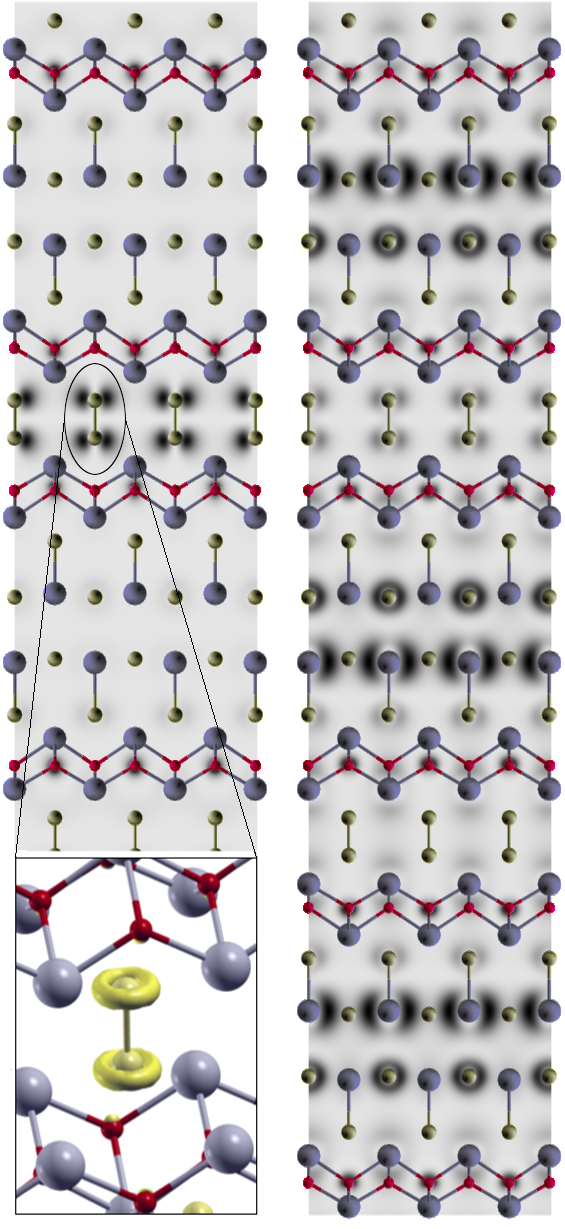}
\caption{Local density of states of Bi$_{3}$O$_{2}$S$_{3}$ in real space integrated in the energy ranges [-1.5,0] (a) and [0,1.5] (b), plotted with XCrySDen \cite{Kokalj2003}. Framed is a 3D plot of an isosurface centred on the S-S dimer. In the first case, the S$_{2}$ layer gathers most of the density. In the second case it is the BiS planes, without the extra S atoms, that gather the density. Half of the atoms represented are not in the plane corresponding to the density plot.}
\label{fig:LDOS}
\end{figure}

To confirm these bands are coming from the S$_{2}$ layer, we plotted the local density of states ("local" refers to decomposing the total density of states into real space), integrated in the range of energies between -1.5 eV and 0 eV  (Figure \ref{fig:LDOS}). The chosen range of energies contains the bands we are interested in.

The results are quite clear: it is indeed the S$_{2}$ layer that gathers all the electron density. Moreover, this density is organised as a torus oriented along the z axis around each S atom (the fact that the density decreases near the atom comes from the pseudopotentials), which shows that the orbitals lying at or just above the Fermi level correspond mainly to \textit{p$_{x}$},\textit{p$_{y}$} electrons on the S$_{2}$ dimers (oriented along z). The concentration of the electron density on the atoms rather than on the S--S bond could be indicative of a weakening of the S--S interaction.

We also plotted the local density of states in real space for the range of energies between 0 eV and 1.5 eV (Figure \ref{fig:LDOS}). This range contains the characteristic bands present in all the BiS$_{2}$-based compounds. We can clearly see that the charges are localised in the BiS$_{2}$ bilayer, more precisely in the two BiS planes. The S atoms in the BiS$_{2}$ bilayer which are outside the BiS planes gather as few electrons in that range of energy as the S atoms in the S$_{2}$ layer.

\subsection{Pinning of the Fermi level}

In the band structures of Bi$_{3}$O$_{2}$S$_{3}$ with and without stacking faults, we observe that the top S$_{2}$ band touches the Fermi level, without crossing it, approximately at equal distance from A and Z. Interestingly, this band is perfectly flat along the $z$ direction: this S$_2$ related band has very marked two-dimensional character, and therefore it shows the characteristic discontinuous 2D Van Hove singularity (very clear at the Fermi level in the three lower panels of Fig. 4), which seems to be pinning the Fermi level. Such a discontinuity is suggestive of possible instabilities along the (1,1,0) direction.

\subsection{Influence of stacking faults}

The electronic structure of Bi$_{3}$O$_{2}$S$_{3}$ with an added stacking fault is very close to that of Bi$_{3}$O$_{2}$S$_{3}$ (Figure \ref{fig:band structures}). The only difference is that some bands below the S$_{2}$ bands are raised slightly. It has strictly no impact on the bands close to the Fermi level.

\section{Conclusions}

The main conclusions of this study are:

\bigskip

\noindent $(i)$ The key result of this study is that, in spite of it being a nominally stoichiometric composition (no doping), the S$_{2}$ layers in the Bi$_{3}$O$_{2}$S$_{3}$ compound act as effective dopants of the parent insulating compound. They push a \textit{p$_{x}$},\textit{p$_{y}$}-like band of the S$_{2}$ layer onto the bottom of the conduction band, thereby transferring part of the electron density from the S$_{2}$ dimers to the BiS planes. This doping is in agreement with experiments, that show that Bi$_{3}$O$_{2}$S$_{3}$ is metallic \cite{Phelan2013}.

Furthermore, the experimental observation of an anomalously short S--S distance in the S$_{2}$ layers is also consistent with electron depletion of the S$_{2}$ layer. Indeed, a substantial residual force is observed for the S atoms in the S$_{2}$ layer in the direction of elongating the distance, which suggests that the real system has a larger depletion of electrons from the S$_{2}$ bands than the one obtained in the calculations.

\bigskip

\noindent $(ii)$ The electronic structure close to the Fermi level is not modified by stacking faults. This indicates that the changes in the superconducting state caused by these are not related to the electronic structure

\bigskip

\noindent $(iii)$ Finally, the calculations of density of states in real space indicate that the BiS bands, shown to be central for superconductivity in these compounds \cite{Usui2012}, actually come from the BiS planes. Therefore finding other structures having such planes would be very interesting in order to tune their interactions differently, and maybe transform their behaviour.

\bigskip

In summary, we studied in detail the electronic structure of the newly discovered superconductor Bi$_{3}$O$_{2}$S$_{3}$. We find the bismuth-sulphur bands corresponding to the BiS planes are doped by the S$_{2}$ dimers, unlike the other superconducting members of the BiS$_{2}$ family which are non-stoichiometric. We studied the influence of stacking faults on this material and found it does not influence the electronic structure. Finally, the fact that the electrons responsible for superconductivity are localised in the 2D bands of BiS planes suggests the possibility of other superconductors with BiS structural units.

\bigskip

\section*{Acknowledgements}

We would like to thank Daniel Sanchez-Portal for his help with particular features of SIESTA, Arman Khojakhmetov for his exploratory work, and Gilbert G. Lonzarich, Richard Needs, Stephen Rowley, Sebastian Haines, Cheng Liu and Adrien Amigues for fruitful discussions. We acknowledge the support from EPSRC, Corpus Christi College, Darwin College, Jesus College, Cambridge Central Asia Forum, Cambridge Kazakhstan Development Trust and KAZATOMPROM. S.E.D would like to thank the Winton Program for the Physics of Sustainability. Part of this work was performed using the Darwin Supercomputer of the University of Cambridge High Performance Computing Service, using Strategic Research Infrastructure Funding from the Higher Education Funding Council for England and funding from the Science and Technology Facilities Council.

\section*{References}

\bibliographystyle{unsrt}
\bibliography{Bi-O-S}

\end{document}